\documentstyle[aps,12pt,preprint,epsfig]{revtex}
\tightenlines

\draft
\begin{document}

\title{Color-singlet direct $J/\psi$ and $\psi'$ production at Tevatron
in the $k_t$ factorization approach}
\author{Feng Yuan}
\address{\small {\it Department of Physics, Peking University, Beijing
100871, People's Republic of China}}
\author{Kuang-Ta Chao}
\address{\small {\it China Center of Advanced Science and Technology
(World Laboratory), Beijing 100080, People's Republic of China\\
and Department of Physics, Peking University, Beijing 100871,
People's Republic of China}}
\maketitle

\begin{abstract}
Direct $J/\psi$ and $\psi'$ production rates at Tevatron are calculated
in the $k_t$-factorization approach within the color-singlet model.
In this approach, the production rates are enhanced by a factor of 20
compared to the naive collinear parton model. However, the theoretical
predictions are still below the experimental data by at least one
order of magnitude.
This means that to explain charmonium productions at Tevatron, we
still need to call for the contributions from color-octet channels
or other production mechanisms.
\end{abstract}

Studies of heavy quarkonium production in high energy collisions provide
important information on both perturbative and nonperturbative QCD.
In recent years, heavy quarkonium production has attracted much attention
from both theory and experiment.
To explain the $J/\psi$ and $\psi'$ surplus problem
of large transverse momentum production at Tevatron\cite{fa},
the color-octet production mechanism was introduced for the
description of heavy quarkonium production\cite{surplus,s1} based
on the NRQCD factorization framework\cite{nrqcd}.
During the last few years, extensive studies have been performed
for the test of this color-octet production mechanism\cite{review}.
However, it was announced recently\cite{teryaev} that the color-octet
contributions to
charmonia productions at Tevatron are excluded by the experimental data
under the $k_t$-factorization approach\cite{ca,co},
where the authors calculated $\chi_{cJ}$
production rates and found that the experimental data
are successfully described only by the color-singlet contributions.
Consequently, an urgent problem arises, which is whether the
direct $J/\psi$ and $\psi'$ productions can also be described
only by the color-singlet contributions in the $k_t$ factorization
approach.
In this paper, we will try to tackle this problem.
We will calculate the color-singlet direct $J/\psi$ and $\psi'$
production in the $k_t$ factorization approach.
We find the production rates
are enhanced by a factor of about 20
compared to the conventional collinear parton model predictions
after considering the $k_t$ effects of the incident partons.
However, the theoretical predictions are still below the
experimental data by at least one order of magnitude.
Therefore, to explain the Tevatron data on large transverse momentum
charmonium productions, we still need to call for color-octet
contributions or other production mechanisms.

The $k_t$-factorization approach differs greatly from the
conventional collinear approximation because it takes the non-vanishing
transverse momenta of the scattering partons into account.
The conventional gluon densities are replaced by unintegrated gluon
distributions which depend on the transverse momentum $k_t$.

The lowest order graphs for heavy quark pair production
in $k_t$ factorization approach are plotted in Fig.~1,
where we just give the typical graphs of \cite{co}. The first two types of
graphs (a) and (b) are conventional ones for the gluon-gluon
fusion processes. The third type of graphs (c)
are needed to preserve gauge invariance.
To calculate heavy quarkonium production in this approach, we
must project the $Q\bar Q$ pair into a particular bound state.
Because the $Q\bar Q$ pairs in (b) and (c) are in color-octet,
these two types of graphs will not contribute to heavy quarkonium
production within the color-singlet model.
In fact in \cite{teryaev}, the productions of color-singlet $P$-wave
charmonium states $\chi_{cJ}$ are calculated, where there are only
type (a) graphs contributing to their productions.

For the spin-triplet $S$-wave charmonium states $J/\psi$ and $\psi'$, in the lowest
order, even type (a) graphs of Fig.~1 can not contribute to their
productions because of $C$ (charge) parity conservation.
So, to produce the spin-triplet color-singlet $S$-wave heavy quarkonium states,
one must go to higher order (in terms of $\alpha_s$) processes where 
an additional gluon must be emitted out.
That is to say, in the lowest order, the spin-triplet $S$-wave quarkonia are only
produced in the $Q\bar Qg$ production processes.
In Fig.~2, we give the typical graphs for $Q\bar Qg$ production
in the $k_t$-factorization approach, which include all contributing
graphs in this order calculations in a gauge invariant formalism\cite{co}.
However, among these graphs only type (a)
graphs of Fig.~2 contribute to $S$-wave
quarkonium productions, because other graphs violate
$C$ parity or color conservations.

To calculate the production amplitude given by the type (a)
graphs of Fig.~2, we make a Sudakov decomposition for
every 4-momenta $k_i$ as
\begin{equation}
k_i=\alpha_i p_1+\beta_i p_2+\vec{k}_{iT},
\end{equation}
where $p_1$ and $p_2$ are the momenta of the incoming hadrons.
In the high energy limit, we have $p_1^2=0$, $p_2^2=0$, and
$2p_1\cdot p_2=s$, where $s$ is the c.m. energy squared.
$\alpha_i$ and $\beta_i$ are the momentum fractions of $p_1$ and $p_2$
respectively.
$k_{iT}$ is the transverse momentum, which satisfies
\begin{equation}
k_{iT}\cdot p_1=0,~~~
k_{iT}\cdot p_2=0.
\end{equation}
For the momenta of the incident gluons $q_1$ and $q_2$, we
have the following decomposition\cite{co},
\begin{equation}
q_1=x_1 p_1+q_{1T},~~~~q_2=x_2 p_2+q_{2T}.
\end{equation}
That is to say, the longitudinal component of $q_1$ ($q_2$)
is only in the direction of light-like vector $p_1$ ($p_2$).

Using the above defined Sudakov variables, we can express the
cross section for $J/\psi$ production from the type (a)
graphs of Fig.~2 in the following form,
\begin{equation}
\label{xs}
d\sigma(p\bar p\rightarrow J/\psi X)=
\frac{1}{64\times 16(2\pi)^4}\frac{d\alpha_\psi }{\alpha_\psi}
\frac{d\alpha_3 }{\alpha_3}d^2p_Td^2q_{1T}d^2q_{2T}\frac{f(x_1;q_{1T}^2)}{q_{1T}^2}
\frac{f(x_2;q_{2T}^2)}{q_{2T}^2}\frac{|A_0(q_{1T},q_{2T}|^2}
{q_{1T}^2q_{2T}^2},
\end{equation}
where $p_T$ is the transverse momentum of $J/\psi$, $\alpha_\psi$ and
$\alpha_3$ are the momenta fractions of $p_1$ carried by $J/\psi$
and the outgoing gluon.
The amplitude $A_0$ describes $J/\psi$ production in the gluon-gluon
fusion processes $g+g\rightarrow J/\psi+g$, with the off-shellness
of the two incident gluons being $q_1^2=-q_{1T}^2$ and
$q_2^2=-q_{2T}^2$ respectively.
We have checked that the amplitude $A_0$ vanishes when
$q_{1T}\rightarrow 0$ or $q_{2T}\rightarrow 0$,
which is required by the gauge invariance of this approach.

$f(x;q_T^2)$ is the unintegrated gluon distribution, which is related to
the conventional gluon distribution by
\begin{equation}
xg(x,\mu^2)=\int^{\mu^2} \frac{dk_T^2}{k_T^2}f(x;k_T^2).
\end{equation}
The unintegrated gluon distribution $f(x;q_T^2)$ includes the evolution
in $x$ and $k_T^2$ by the BFKL and DGLAP equations. In the nonperturbative
region of small $k_T^2$ the unintegrated gluon distribution is not known,
so in practice the above equation may be rewritten as\cite{co,kwi,ryskin}
\begin{equation}
xg(x,\mu^2)=xg(x,q_0^2)+\int_{q_0^2}^{\mu^2} \frac{dk_T^2}{k_T^2}f(x;k_T^2),
\end{equation}
which introduces a priori unknown initial scale $q_0^2$ and the
initial gluon distribution $xg(x,q_0^2)$.
According to this procedure, the amplitude has the following
decomposition form\cite{co},
\begin{eqnarray}
\nonumber
S(q_{1T},q_{2T})&=&\frac{|A_0(q_{1T},q_{2T})|^2}{q_{1T}^2q_{2T}^2}\\
\nonumber
&=&S(0,0)\theta(q_0^2-q_{1T}^2)\theta(q_0^2-q_{2T}^2)+
S(q_{1T},0)\theta(q_{1T}^2-q_0^2)\theta(q_0^2-q_{2T}^2)\\
&+&S(0,q_{2T})\theta(q_0^2-q_{1T}^2)\theta(q_{2T}^2-q_0^2)+
S(q_{1T},q_{2T})\theta(q_{1T}^2-q_0^2)\theta(q_{2T}^2-q_0^2).
\end{eqnarray}
The function $S$ here is similar to the function $I$ of \cite{co} called
the impact factor, which is, in effect, an off-shell but gauge
invariant cross section\cite{co}.
We note that $S(0,0)$ represents the cross section
for gluon-gluon fusion processes with no $k_t$ effects associated
with the two incident gluons. Substituting this part of $S$ into
Eq.~(\ref{xs}), we can reproduce the conventional collinear parton model
results for $S$-wave quarkonium production.

For numerical calculations, we choose the unintegrated gluon distribution
of \cite{martin}, where the authors determined it by using a combination
of DGLAP and BFKL equations. With the initial conditions
$$
q_0^2=1GeV^2,~~xg(x, q_0^2)=1.57(1-x)^{2.5},
$$
they obtained an excellent fit to $F_2(x,Q^2)$ data over a
wide range of $x$ and $Q^2$.
As in Refs.\cite{levin,teryaev}, we set the scales $\mu^2$ for
strong coupling constant $\alpha_s(\mu^2)$ in the
amplitude squared $|A_0|^2$ to be $q_{1T}^2$ for the interaction
vertex associated with the incident gluon $q_1$, $q_{2T}^2$ for
the vertex associated with $q_2$, and $p_T^2+m^2$
($m=2m_c$ is the quarkonium mass)
for the vertex associated with the outgoing gluon
attached to the $c\bar c$ line.

In the following we present the numerical results for color-singlet direct
$S$-wave charmonia productions
at Tevatron, compared with the experimental data\cite{fa}. Fig.~3 is for
$J/\psi$, and Fig.~4 is for $\psi'$. In the calculations, we use
charm quark mass as $m_c=1.5GeV$,
and the color-singlet matrix elements as
$\langle {\cal O}_1^\psi({}^3S_1)\rangle =1.08 GeV^3$,
and $\langle {\cal O}_1^{\psi'}({}^3S_1)\rangle =0.67 GeV^3$\cite{bodwin}.
To compare with the experimental data we have imposed a cut on
the rapidity of $J/\psi$ and $\psi'$: $|\eta|<0.6$.
From these two figures, we can see that the 
production rates are enhanced by a factor of about 20
compared to the conventional collinear parton model predictions
after considering the $k_t$ effects of the incident gluons.
However, the theoretical predictions are still below the
experimental data by at least one order of magnitude.
Therefore, to explain the Tevatron data on large transverse momentum
charmonium productions, we still need to call for color-octet
contributions or other production mechanisms.

In conclusion, we have calculated direct $J/\psi$ and $\psi'$ production
at Tevatron in 
$k_t$-factorization approach within the color-singlet model.
We found that the production rates are enhanced by a factor of 20
compared to the conventional collinear parton model.
However, the theoretical predictions are still below the
experimental data by at least one order of magnitude.
This means that to explain charmonium productions at Tevatron, we
still need to call for the contributions from color-octet channels
or other production mechanisms.

\acknowledgments
This work was supported in part by the National Natural Science Foundation
of China, the State Education Commission of China, and the State
Commission of Science and Technology of China.

\newpage
\newpage
\vskip 10mm
\centerline{\bf \large Figure Captions}
\vskip 1cm
\noindent
FIG. 1. The typical graphs for $Q\bar Q$ production in $k_t$ factorization
approach in gauge invariant formalism.

\noindent
FIG. 2. The typical graphs for $Q\bar Qg$ production in $k_t$ factorization
approach in gauge invariant formalism.

\noindent
FIG. 3. The color-singlet predictions for large $p_T$ direct $J/\psi$
production compared with the experimental data from Tevatron.
The dashed line denotes the result in the conventional collinear
parton model, and the solid line is the result in the $k_t$
factorization approach.

\noindent
FIG. 4. The same as Fig.~3 but for $\psi'$.

\begin{figure}[thb]
\begin{center}
\epsfig{file=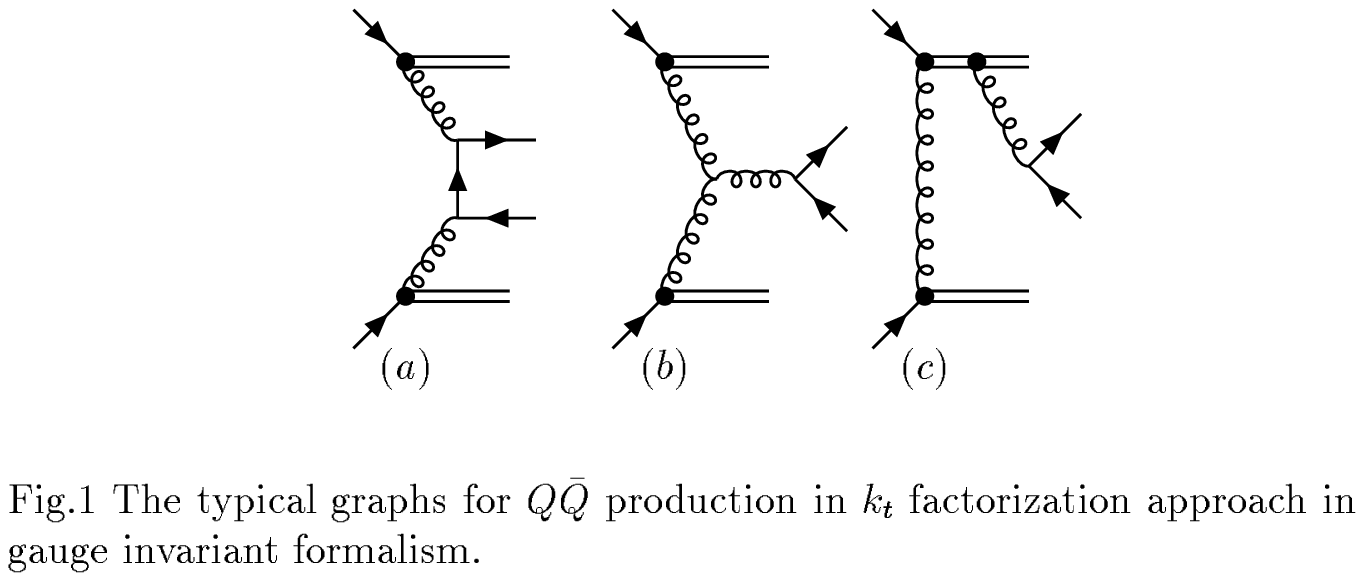,angle=0,width=16cm}
\end{center}
\end{figure}

\begin{figure}[thb]
\begin{center}
\epsfig{file=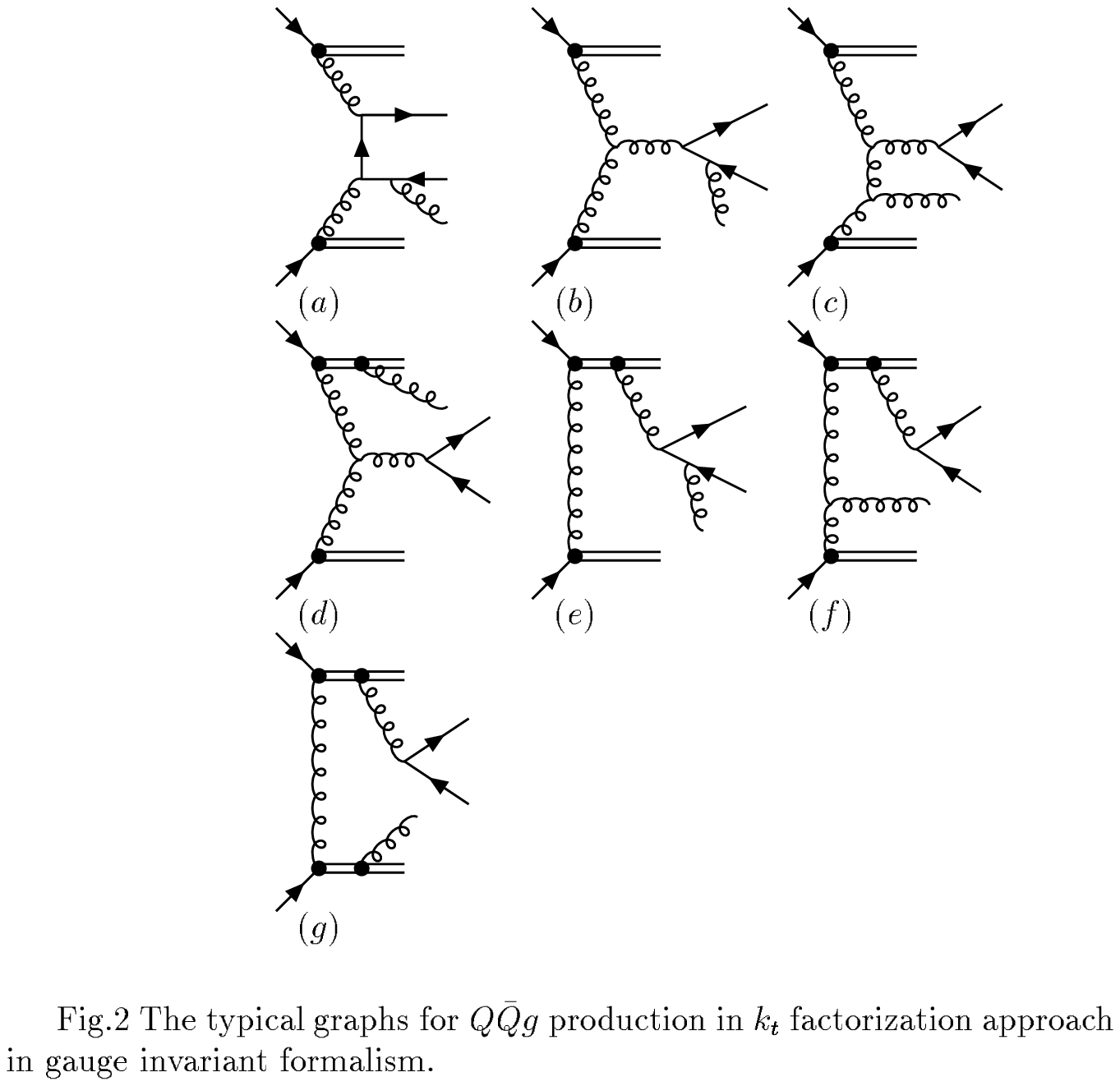,angle=0,width=16cm}
\end{center}
\end{figure}

\begin{figure}[thb]
\begin{center}
\epsfig{file=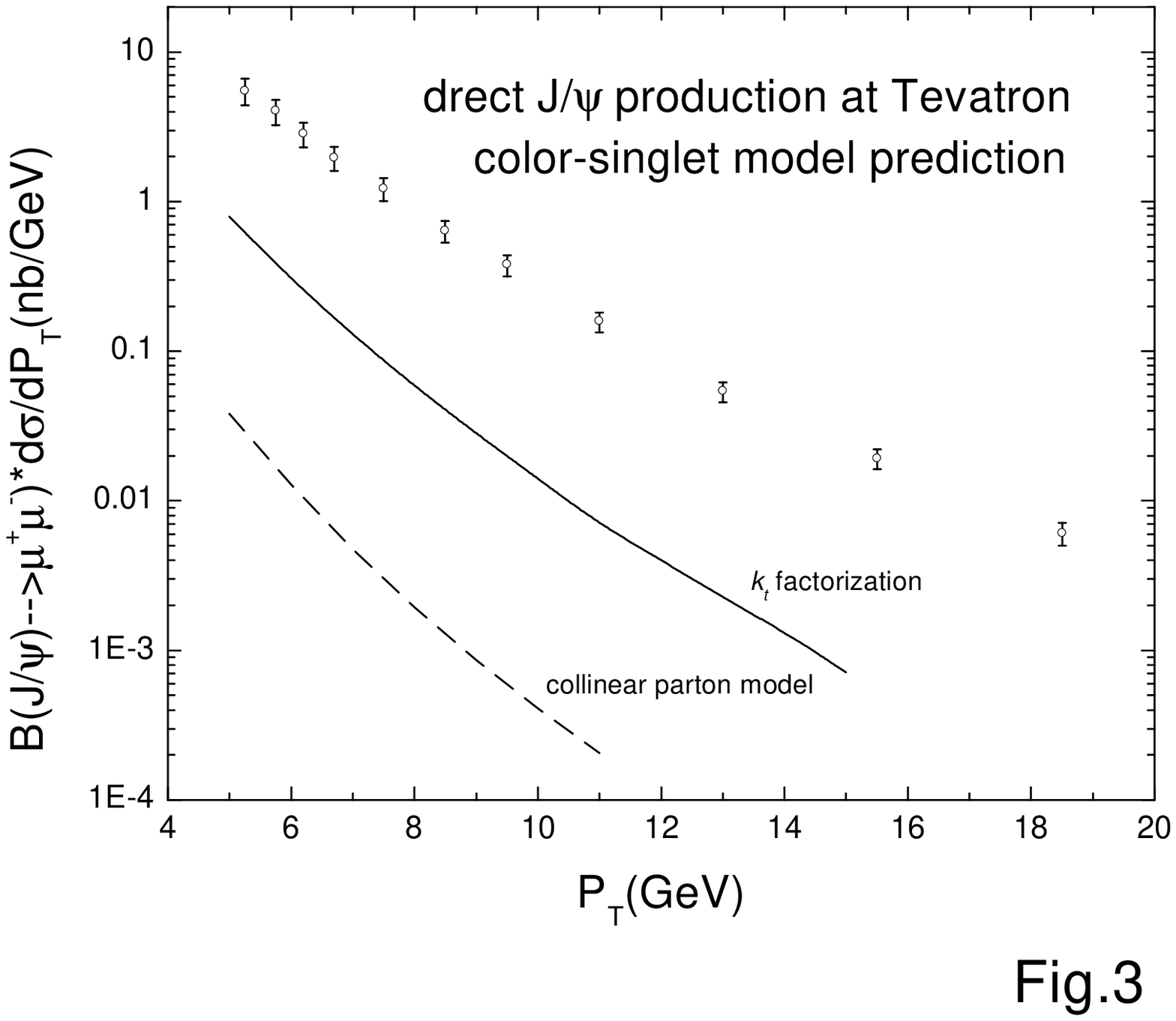,angle=0,width=16cm}
\end{center}
\end{figure}

\begin{figure}[thb]
\begin{center}
\epsfig{file=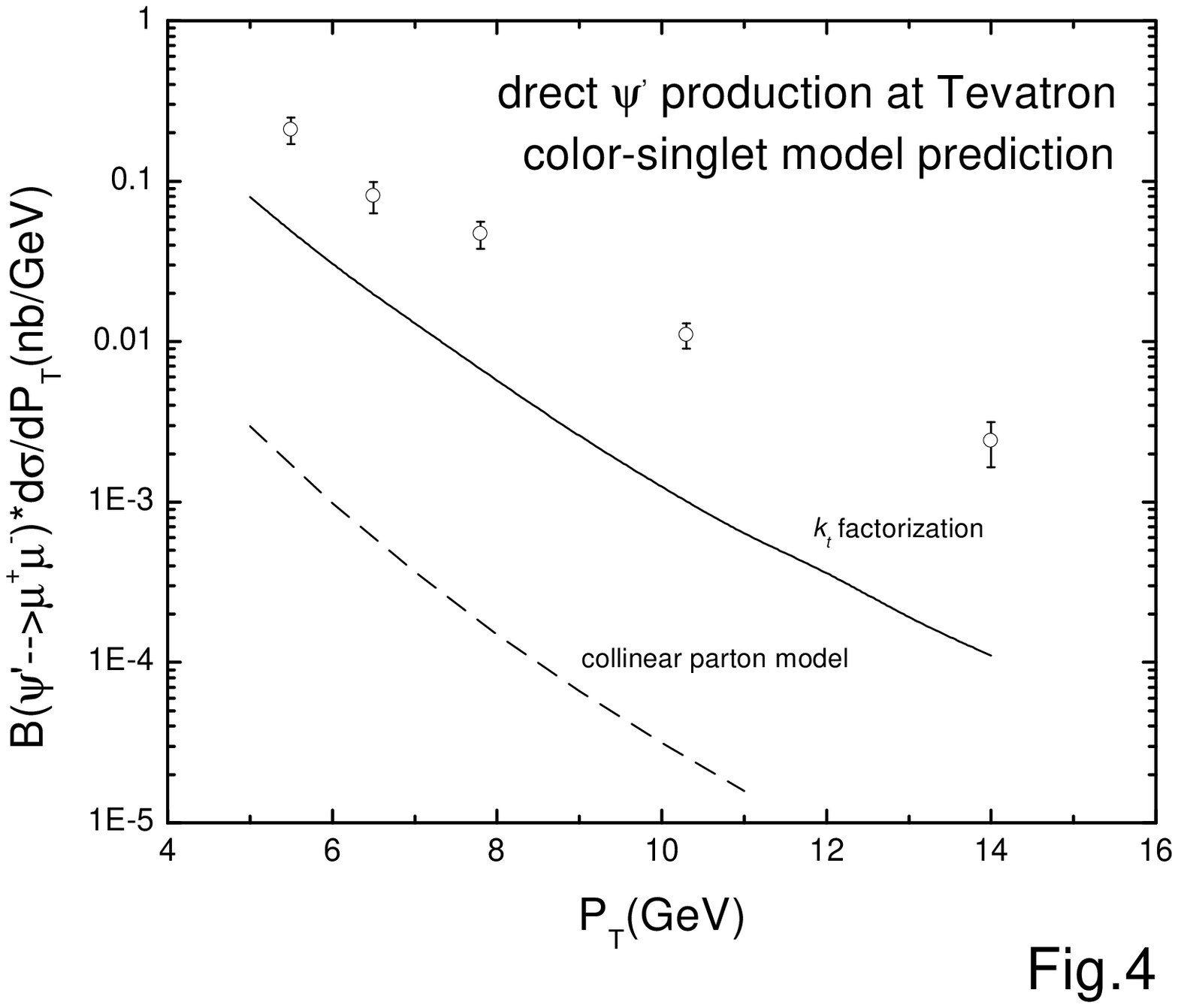,angle=0,width=16cm}
\end{center}
\end{figure}

\end{document}